\documentclass[11pt, a4paper]{article}

\pdfoutput=1

\usepackage[thicklines]{cancel}
\usepackage{slashed}
\usepackage{jheppub}
\usepackage{graphicx}
\usepackage{amssymb}
\usepackage{slashed}
\usepackage{amsmath}
\usepackage{url}
\usepackage{comment}
\usepackage[sort&compress]{natbib}
\usepackage{hyperref} 

\hypersetup{
    colorlinks=true,       
    linkcolor=red,          
    citecolor=green,        
    filecolor=magenta,      
    urlcolor=cyan           
}

\newcommand{\eref}[1]{\ensuremath{\mathrm{Eq.}\;(\ref{#1})}}

\newcommand{\GeV}{\ \mathrm{GeV}}

\newcommand{\Cwronggm}{green}
\newcommand{\Cnopt}{orange}
\newcommand{\Csfopt}{blue}
\newcommand{\Ctwopt}{gray}
\newcommand{\Cesp}{purple}
\newcommand{\Ctachyon}{red}

\title{Strongly First Order Phase Transitions Near an Enhanced
  Discrete Symmetry Point}

\author[a]{Vernon Barger,}
\author[a]{Daniel J. H. Chung,}
\author[a]{Andrew J. Long,}
\author[b]{and Lian-Tao Wang}

\affiliation[a]{Department of Physics, University of Wisconsin, Madison, WI 53706}
\affiliation[b]{Enrico Fermi Institute and Department of Physics, University of Chicago, Chicago, IL 60637}

\emailAdd{barger@pheno.wisc.edu}
\emailAdd{danielchung@wisc.edu}
\emailAdd{ajlong@wisc.edu}
\emailAdd{liantaow@uchicago.edu}
\abstract{
We propose a group theoretic condition which may be applied to extensions of the Standard Model in order to locate regions of parameter space in which the electroweak phase transition is strongly first order, such that electroweak baryogenesis may be a viable mechanism for generating the baryon asymmetry of the universe.  
Specifically, we demonstrate that the viable corners of parameter space may be identified by their proximity to an enhanced discrete symmetry point.  
At this point, the global symmetry group of the theory is extended by a discrete group under which the scalar sector is non-trivially charged, and the discrete symmetry is spontaneously broken such that the discrete symmetry relates degenerate electroweak preserving and breaking vacua.  This idea is used to investigate several specific models of the electroweak symmetry breaking sector.  
The phase transitions identified through this method suggest implications for other relics such as dark matter and gravitational waves.
}


\begin{document}
\maketitle


\section{Introduction}

Standard cosmology of the early universe within the context of a large class of models embedding the Standard Model (SM) of particle physics predicts the existence of an electroweak symmetry breaking (EWSB) phase transition (PT).  
Collider constraints alone cannot determine the nature of the EWSB PT in a model independent way.  
However, additional information is available in the form of cosmological relics, which were produced in the early universe and survive as direct probes of the physics of the era during which the temperature was electroweak scale.  
Relics such as the baryon asymmetry \cite{Kuzmin:1985mm}, primordial gravitational waves \cite{Chung:2010cb, Huber:2008hg,Caprini:2009yp,Kahniashvili:2009mf}, and (modifications to) the dark matter relic abundance \cite{Chung:2011it, Chung:2011hv, Wainwright:2009mq, Cohen:2008nb}, may have been generated at the electroweak scale PT(s).  

Generating the baryon asymmetry through CP violations at electroweak symmetry breaking bubbles \cite{Kuzmin:1985mm}, requires a strongly first order phase transition (SFOPT) to protect the baryon number in the broken phase. 
In this context, a SFOPT may be defined as a first order PT in which the (thermal) expectation value of the SM-like Higgs $v(T) = \left< h \right>$ satisfies $v(T) / T \gtrsim 1$ in the broken phase after the phase transition completes,
such that weak sphaleron processes are inactive \cite{Klinkhamer:1984di,Kuzmin:1985mm}.  
It is well-known that the SM is unable to accommodate a SFOPT while satisfying the Large Electron-Positron (LEP) Collider bounds on the Higgs mass \cite{Csikor:1998eu}.  
This is one of the main motivations for considering an extended Higgs sector. 
Many beyond the Standard Model theories are able to accommodate a SFOPT, including supersymmetry, two Higgs doublet models, and minimal scalar singlet extensions of the SM.  
However, if the extra scalar fields obtain vacuum expectation values (vevs), one often finds that new patterns of symmetry breaking become accessible.  
This fact makes the phase transition more difficult to study, because quantities such as $v(T) / T$ are nonanalytic functions of the parameters of the model.  
Consequently, many beyond the Standard Model PT analyses rely on an intensive numerical parameter scan to search for SFOPT.  
Although such scans may be capable of locating SFOPTs, on their own they do not reveal why one particular parametric limit is favored over another.

In this article, we propose a group theoretic guideline which will aid the search for SFOPT in a large parameter space and help to identify why certain parametric limits are favored over others.  
Our guideline is motivated by the following heuristic argument.  
In perturbative thermal effective potential computations, the thermal mass is of the order $c \, T^2$ where $c$ is a thermal loop factor. 
Therefore, if all the renormalized coupling constants are of order unity and all mass scales are of the electroweak scale, we expect that the phase transition will occur at a temperature $T \sim v/\sqrt{c}$ such that $v(T)/T \sim \sqrt{c} < 1$, and the PT is typically not strongly first order.  
Hence, in order to have a SFOPT, the renormalized parameters of the theory must be near a special point in the parameter space.
An ideal parametric limit which overcomes the natural thermal loop suppression is the region where $v(T)/T \rightarrow \infty$.  
To achieve this, it would be unnatural to expect $v(T)$ to deviate by many orders of magnitude from the electroweak scale, because of the constraint that $v(0)$ defines the electroweak scale. 
On the other hand, $v(T) / T$ may be enhanced by taking the $T \to 0$ limit.  

The limit of low phase transition temperature and large $v(T) / T$ can be achieved naturally by employing a discrete symmetry.  
The phase transition begins at the critical temperature $T_c$, defined as the temperature above which the thermal corrections are sufficiently large as to make the EW symmetric phase energetically favored, and below which the EW broken phase is favored.  
Hence, at $T = T_c$ the thermal effective potential possesses two degenerate minima corresponding to the EW symmetric and broken phases (see also Appendix \ref{app:tempdefs}).  
One may enhance $v(T_c) / T_c$ by taking $T_c \to 0$ provided that there is a mechanism guaranteeing that the theory possesses such degenerate vacua even in the absence of thermal corrections.  
One mechanism that yields degenerate vacua is the spontaneous symmetry breaking of a discrete group (see e.g.~\cite{Kibble:1976sj,vilenkin2000cosmic}).  
After spontaneous symmetry breaking, one finds a set of degenerate vacuum states which fall into a coset representation of the discrete group.  
Moreover, if the discrete symmetry group does not commute with the electroweak group, then the scenario described above may be achieved: the electroweak symmetry is broken in one vacuum and unbroken in a second degenerate vacuum implying $T_c = 0$ and $v(T_c) / T_c = \infty$.  

Of course the existence of degenerate vacua alone does not imply $v(T) / T \gg 1$, since the EW phase transition must take place, and this is not necessarily the case in extensions of the SM with multiple vacua.  
If the discrete symmetry is exact, then $T_c = 0$ and the phase transition does not proceed because the broken phase never becomes energetically favored.  
Hence, we will consider models in which the discrete symmetry is generally approximate, but becomes exact at a particular parametric point, referred to as an enhanced discrete symmetry point (EDSP)\footnote{In general, a model may possess multiple EDSPs each relating the EW broken and symmetric vacua by a different symmetry transformation.}.    
Then the heuristic arguments above imply that one can expect to find SFOPT in a parametric neighborhood of an EDSP and connected to it by a continuous ``small'' deformation which breaks the discrete symmetry.  
Precisely how ``small'' a deformation is required depends upon two model-dependent conditions: the condition that the electroweak PT must complete and upon the order unity number that sits at the right of the inequality $v(T)/T > 1$.  
Hence the takeaway message is that one can make the analysis of and search for SFOPT in a large parameter space more tractable with the aid of an EDSP ``lamppost" which signals the parametric neighborhood which is favorable for SFOPT.

The order of presentation is as follows.  In Sec.~\ref{sec:whydiscrete} we motivate our group theoretic identification of SFOPTs.  In Sec.~\ref{sec:examples}, we employ our  technique to explore three example models.  
We then finish with some concluding remarks in Sec.~\ref{sec:conc} and an appendix which reviews some relevant basics of phase transitions used in this paper.


\section{\label{sec:whydiscrete} Why Discrete Symmetry?}
Suppose that a given theory is exactly invariant under an internal discrete symmetry group $G$.  
It is well-known that the spontaneous symmetry breaking of $G$ down to $H \subset G$ leads to the vacua giving a nontrivial coset $G/H$ representation \cite{Kibble:1976sj,vilenkin2000cosmic}.  
We will first illustrate how this connects to a SFOPT in a perturbative single real scalar field toy model, and then proceed to give a more general discussion.  

Quite often in extensions of the SM, other scalar fields along with the Higgs obtain vevs at the electroweak phase transition.  
One may model such a first order phase transition with the following toy theory in which $\varphi$ represents the linear combination of the SM Higgs and other scalar fields.  
Consider the theory of a real scalar field $\varphi$  with the classical potential
\begin{align}\label{eq:GenSing_U}
	U(\varphi) = \frac{1}{2} M^2 \varphi^2 - \mathcal{E} \varphi^3 + \frac{\lambda}{4} \varphi^4  \, ,
\end{align}
and suppose that $\varphi$ is coupled to a family of $N$ fermions $\mathcal{L} \supset (m_i + h_i \varphi) \bar{\psi}_i \psi_i$.  
Note that this theory has no internal symmetries for non-special values of the parameters $\{M^2, \mathcal{E},\lambda, h_i, m_i\}$.  
When we turn on temperature, there will be a thermal bath of $\varphi$ and $\psi_i$ particles.  
If the fermions are relativistic at the electroweak scale (i.e., $m_i^2 \ll T^2$), then the thermal effective potential can be written to leading order as
\begin{equation}
V_{\rm eff}(\varphi, T) \approx U(\varphi) + c \, T^2 \varphi^2
\label{eq:leadingapproxtoy}
\end{equation} 
where $c \approx N h_i^2 / 12$ \cite{Kapusta:1989}.  
Here, in the so-called high-temperature approximation, we have neglected the subdominant thermal corrections (such as the non-analytic term) and the $\hbar$ radiative corrections.

As long as the supercooling is small (e.g., as measured by the fractional temperature change during the duration of the PT), the PT occurs at the temperature near $T_c$ at which the thermal effective potential $V_{\rm eff}$ displays two degenerate minima (for more details, see Sec. \ref{app:tempdefs}).  
Solving this constraint for $T_c$ gives
\begin{align}\label{eq:GenSing_Tc}
	T_c^2 = \frac{\mathcal{E}^2}{\lambda \, c} \left( 1 - \frac{\lambda M^2}{2 \mathcal{E}^2} \right) \, .
\end{align}
In this simple toy model and subject to the approximation \eref{eq:leadingapproxtoy}, there is an enhanced $\mathbb{Z}_2$ symmetry at $T=T_c$.  
Explicitly, the potential at $T=T_c$ becomes
\begin{equation}\label{eq:VatTc}
	V_{\rm eff}(\varphi, T_c)= \frac{\lambda \varphi^2}{4} \left( \varphi -\frac{2 \mathcal{E}}{\lambda} \right)^2 \, 
\end{equation}
and respects the discrete symmetry
\begin{equation}
\label{eq:symmetry}
	\mathbb{Z}_2 \ : \ \left( \varphi - \frac{\mathcal{E}}{\lambda} \right) \to - \left( \varphi - \frac{\mathcal{E}}{\lambda} \right) \, ,
\end{equation}
which was not originally present in \eref{eq:GenSing_U}.  
This $\mathbb{Z}_2$ exchanges the degenerate vacua at $\varphi = 0$ and $\varphi = v(T_c) = 2 \mathcal{E} / \lambda$ across a potential barrier at $\varphi = \mathcal{E} / \lambda$.  
It is to be noted that \eref{eq:VatTc} is independent of $M$, and thus this symmetry exists in this toy model for any critical temperature $T_c$ that can be tuned using $M$.\footnote{Although this is an enhanced $\mathbb{Z}_2$ symmetry at $T = T_c$, since the symmetry does not generically exist at other temperatures, it is not the enhanced discrete symmetry point relevant for this paper.  See below for further clarification.  }

Although there is no electroweak symmetry in this toy model, there is still a first order phase transition, and we can investigate the parametric dependence of its order parameter $v(T_c) / T_c$.  
Since $v(T_c) = 2 \mathcal{E} / \lambda$ is independent of $M^2$, the order parameter can be maximized by varying $M^2$ to minimize $T_c$.  
Even though the high-temperature expansion breaks down when $T$ drops below the mass of the fermion, the formal limit $T_c \rightarrow 0$ can be taken assuming that the fermions are massless.  
The formal solution to $T_c=0$ is\footnote{ The idea of focusing on $T_c \approx 0$ was recently emphasized by \cite{Espinosa:2011ax}.}  
\begin{equation}
	\alpha \equiv \lambda M^2 / 2 \mathcal{E}^2=1 \, .
\end{equation}
The important observation is that $1 - \alpha = 0$ corresponds to an EDSP in the parameter space at which the {\em zero-temperature} scalar potential \eref{eq:GenSing_U} is invariant under the symmetry transformation Eq.~(\ref{eq:symmetry}).  
At the EDSP, $1 - \alpha = 0$, the order parameter 
\begin{align}
	\frac{v(T_c)}{T_c} = 2 \sqrt{\frac{c}{\lambda}} \frac{1}{\sqrt{1 - \alpha}} 
\end{align}
formally diverges, and for $1 - \alpha \ll 1$, the phase transition may be made arbitrarily strongly first order.  

Hence, our group theoretic guideline leads us to identify the parametric region in the vicinity of the EDSP $1 - \alpha = 0$ as favorable for SFOPT.  
However, for this region to be truly viable, it must be the case that the rate at which bubbles of the broken phase nucleate is sufficiently large that the phase transition actually completes.  
This requires the discrete symmetry to be weakly broken, such that the PT occurs at a nonzero $T$.\footnote{Note that the bubble nucleation rate is zero at $T=T_c$.}  
In the toy model, such breaking can be accomplished explicitly at the classical level through a finite excursion from the EDSP (i.e.  $1 - \alpha = \epsilon \neq 0$), or radiatively through the Yukawa coupling.  
Indeed, in many extensions of the SM where singlets are introduced, the relevant discrete symmetry transforms both the Higgs and the singlet fields.  
Since the singlets lack SM gauge couplings, radiative corrections necessarily break the discrete symmetry to a degree controlled by the strength of the gauge interactions.  
If the breaking of the symmetry is so large that the potential does not have the qualitative features of \eref{eq:GenSing_U} near $\alpha = 1$, then the EDSP method loses its advantage for identifying SFOPT.  
If the breaking of the symmetry is so small that bubbles will not nucleate fast enough to complete the PT, then any candidate parameter points found with the EDSP method are inherently not viable.  
Since this non-completion of the PT will be a general feature of the region of parameter space nearby to the EDSP, we must take extra care in choosing the temperature at which to evaluate the EW order parameter $v(T) / T$.  
Up to this point in the discussion, we have evaluated $v(T) / T$ at the degeneracy temperature $T_c$, which, physically, is the temperature in the symmetric phase at the onset of supercooling.  
However, a first order phase transition proceeds with the nucleation of bubbles of broken phase which subsequently collide and reheat the plasma to a temperature $T_r$ (see App.~\ref{app:tempdefs} for precise definition).  
Since the purpose of the SFOPT criterion $v(T) / T \gtrsim 1$ is to ensure suppression of weak sphaleron processes {\it in the broken phase after the phase transition}, the most physically relevant temperature at which to evaluate $v(T) / T$ is the reheat temperature $T_r$.  

\begin{figure}[t]
\begin{center}
\includegraphics[width=0.3\textwidth]{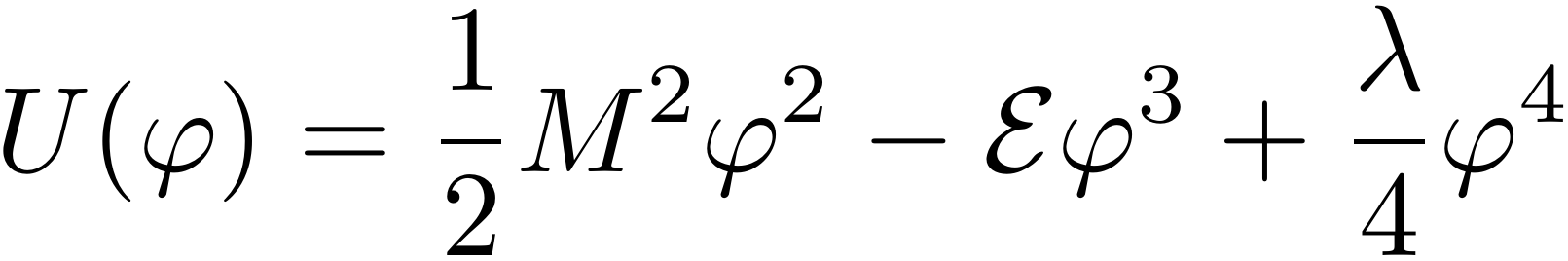}  \\
\includegraphics[width=0.8\textwidth]{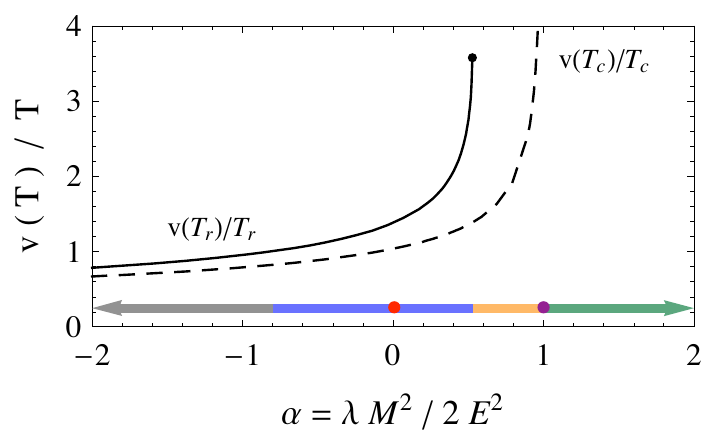}  \\
\includegraphics[width=0.18\textwidth]{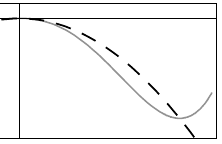} \includegraphics[width=0.18\textwidth]{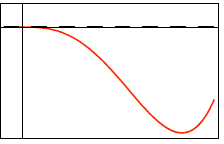} 
  \includegraphics[width=0.18\textwidth]{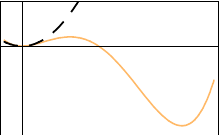} \includegraphics[width=0.18\textwidth]{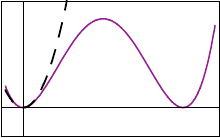} \includegraphics[width=0.18\textwidth]{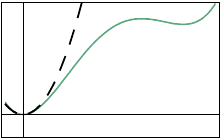}
\caption{\label{fig:GenSing_scan1}  The order parameter, calculated as $v(T_c) / T_c$ (dashed) and $v(T_r) / T_r$ (solid), plotted against $\alpha = \lambda M^2 / 2 \mathcal{E}^2$.  The insets show $U(\varphi)$ for particular values of $\alpha$ in each of the associated colored regions.  
}
\end{center}
\end{figure}

To obtain a numerical intuition for our proposal, consider Fig.~\ref{fig:GenSing_scan1} where we have plotted $v(T_c) / T_c$ (dashed) and $v(T_r) / T_r$ (solid) while varying $\alpha$ and fixing $U^{\prime}(v) = 0$ at $v = 300 \GeV$, $U^{\prime \prime}(v) = (50 \GeV)^2$, $N = 1$, and $h=0.3$.  
In this figure, we also show $U(\varphi)$, such as to make the discrete symmetry evident at the EDSP.  
As expected, $v(T_c) / T_c$ diverges at the EDSP and is arbitrarily large for arbitrarily small discrete symmetry breaking ($1 - \alpha \ll 1$).  
On the other hand, $v(T_r) / T_r$ cannot be calculated if the discrete symmetry is too weakly broken ($1 - \alpha \lesssim 0.5$), because the phase transition does not occur.  
However, sufficient discrete symmetry breaking ($1 - \alpha \gtrsim 0.5$) yields SFOPT which become monotonically weaker as the degree of symmetry breaking grows.  
We have used the same coloring in Fig.~\ref{fig:GenSing_scan1} as we do in the rest of this article to distinguish the varous regions of parameter space: the phase transition does not occur because the broken phase is not energetically favored (\Cwronggm); the PT does not occur because the bubble nucleation rate is too low (\Cnopt); a strongly first order PT occurs (\Csfopt); a weakly first order or second order PT occurs (\Ctwopt); the EDSP (\Cesp \ dot); and the point at which the barrier disappears (\Ctachyon \ dot).

Now let us return to a more broad discussion of the connection between discrete symmetry and strongly first order phase transition.  
In retrospect, we recognize that the existence of an EDSP associated with a discrete symmetry under which the vacua form a coset representation (along with the condition that spontaneous symmetry breaking occurs) is sufficient to obtain $v(T_c)/T_c \rightarrow \infty$ since $T_c=0$ implies a degeneracy at the level of the nonthermal effective potential.
Even though the toy model calculation was accomplished using the leading high-temperature $T$ dependence and the classical potential, this statement regarding the EDSP is an exact statement for an exact effective potential.  
In other words, as far as this exact statement is concerned, it is not particularly important that $T=T_c$ corresponded to an enhanced symmetry point for general $T_c$ as in the case of this simple one dimensional toy model (see \eref{eq:VatTc}), nor is it important that quantum radiative corrections from the Yukawa couplings break the discrete symmetry given by Eq.~(\ref{eq:symmetry}).
One final ingredient, which is important for electroweak baryogenesis but is not represented in the toy model is that at least two vacua in the coset space must carry different electroweak quantum numbers.
Otherwise, the PT will not be an electroweak symmetry breaking PT.
This means that the discrete group must not commute with the electroweak group and one element in the coset representation must be an electroweak singlet.  
Hence our group theoretic guideline may be summarized as:  an arbitrarily strong phase transition (i.e., $v(T_c) / T_c \gg 1$) may be found in the parametric neighborhood of an EDSP if 1) the condition for spontaneous (discrete) symmetry breaking is satisfied (such that there will be degenerate vacua), 2) the discrete group does not commute with the electroweak group, and 3) its coset representation contains an electroweak singlet element (such that the EW symmetry is broken in one vacuum and preserved in another).


\vspace{1cm}
\section{\label{sec:examples}A Few Examples}


\subsection{SM with Low Cutoff}\label{sub:NRSM}

\begin{figure}[t]
\begin{center}
\includegraphics[width=0.60\textwidth]{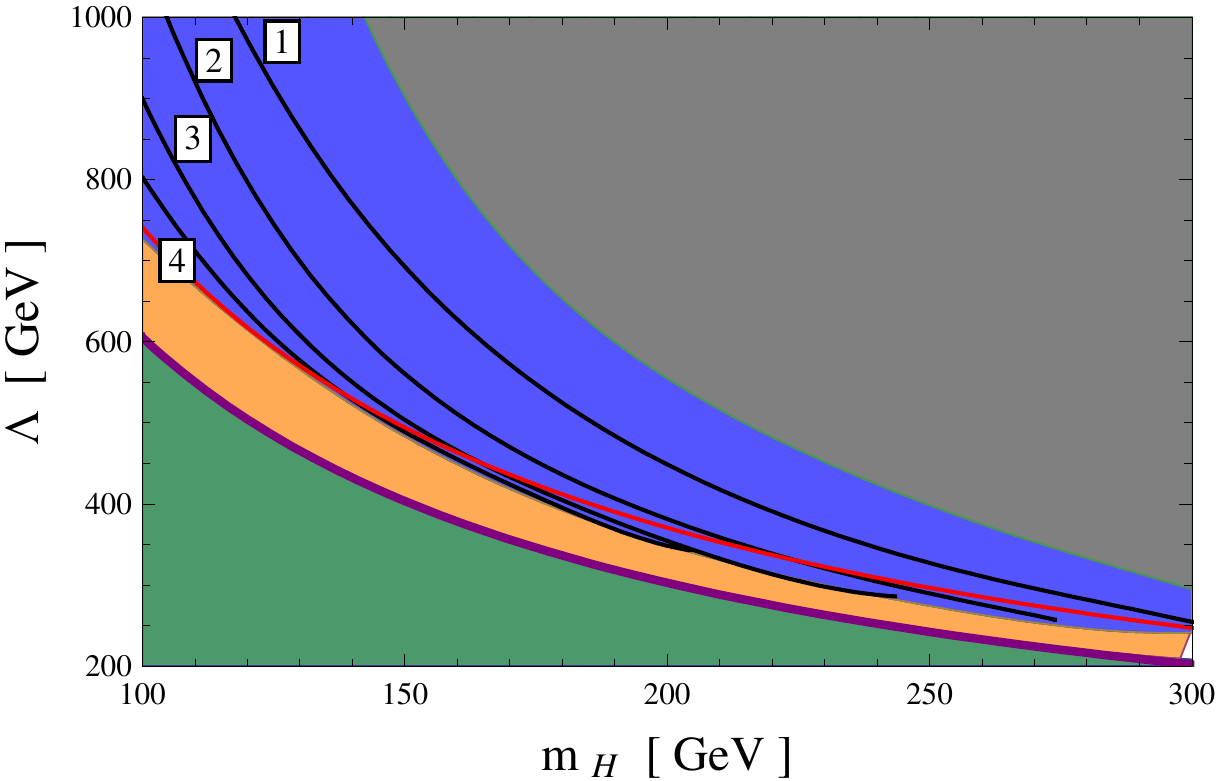}
\caption{\label{fig:NRSM_scan1}  
The parameter space nearby to the EDSP (\Cesp \ curve).  The coloring is the same as in Fig.~\ref{fig:GenSing_scan1}.  The PT order parameter $v(T_r)/T_r$ is indicated by the overlaid contours.  SFOPTs are found in the \Csfopt \  region and become weaker in the \Ctwopt \ region, farther from the EDSP.  
}
\end{center}
\end{figure}

As a first example, we will consider a generic extension of the SM with a low scale cutoff, as studied by \cite{Grojean:2004xa, Delaunay:2007wb, Barger:2003rs}.  
Provided that the UV physics does not violate the EW symmetry, then upon integrating it out one obtains a classical potential of the form
\begin{align}\label{eq:LowCut_U}
	- \mathcal{L} \supset \lambda \left( \bigl| H^{\dagger} H \bigr| - \frac{v^2}{2} \right)^2 + \frac{1}{\Lambda^2} \left( \bigl| H^{\dagger} H \bigr| - \frac{v^2}{2} \right)^3
\end{align}
up to terms of order $H^8 / \Lambda^4$.  
Writing the Higgs doublet in terms of the fundamental scalar Higgs $h$ as $H = \left( 0 , h / \sqrt{2} \right)^{\rm T}$, and using $m_H^2 = 2 \lambda v^2$, the potential becomes
\begin{align}
	U(h) = \frac{1}{8 \Lambda^2} h^6 - \frac{\lambda}{4} \left( 3\frac{v^4}{m_H^2 \Lambda^2}  - 1 \right) h^4 + \frac{\lambda v^2}{4} \left( 3\frac{v^4}{m_H^2 \Lambda^2} -2 \right) h^2 
\end{align}
up to constant and higher order terms.  
There exists an enhanced discrete symmetry point\footnote{It may be more appropriate to use the term ``enhanced discrete symmetry plane,'' as the condition $m_H \Lambda = v^2$ actually specifies a hypersurface in the parameter space, but we will continue using EDSP for simplicity.  } at which a $\mathbb{Z}_2$ symmetry is nonlinearly realized,
\begin{align}
	{\rm EDSP:} \quad m_H \Lambda = v^2
	\qquad \qquad \qquad
	\mathbb{Z}_2^{\prime} \ : \ h \to - \frac{h}{2} + \sqrt{v^2 - \frac{3}{4} h^2} \, .
\end{align}
The $\mathbb{Z}_2^{\prime}$ symmetry exchanges the minima at $ h = 0$ and $h = v$ while leaving the maximum at $h = v / \sqrt{3}$ invariant.  
We have reproduced an earlier PT analysis \cite{Grojean:2004xa} in order to illustrate the proximity of SFOPTs to the EDSP.  
Moreover, we have extended the previous analysis by calculating the more physically relevant order parameter $v(T_r) / T_r$, instead of $v(T_c) / T_c$.  
Our results are summarized in Fig.~\ref{fig:NRSM_scan1}, and are in good agreement with Fig.~2 of \cite{Grojean:2004xa} which shows the same slice of parameter space.  
We find that nearby to the EDSP (\Cesp \ curve), the PT is strongly
first order (\Csfopt), and that the PT becomes weaker moving away from the EDSP.  
It is also worth noting that while the barrier persists, the PT most likely does not occur, as evidenced by the lack of \Csfopt \ in the region between the \Cesp \ and \Ctachyon \ curves except for a small sliver above $m_H = 200 \GeV$.

\vspace{1cm} 
\subsection{SM Plus Real Singlet -- xSM}\label{sub:xSM}

Next, we will consider models with multiple scalars in the electroweak sector.  
Extending the SM by a real scalar singlet $s$, we obtain a model known as the xSM \cite{Barger:2007im}, which has the classical potential
\begin{align}\label{eq:xSM_U}
	U(h, s) = \frac{\lambda_0}{4} h^4 - \frac{\mu^2}{2} h^2 + \frac{b_4}{4} s^4 + \frac{b_3}{3} s^3 + \frac{b_2}{2} s^2 + \frac{a_2}{4} \, s^2 h^2 + \frac{a_1}{4} s h^2  \, .
\end{align}
Since there is no symmetry protecting $s = 0$, generally both $h$ and $s$ will obtain vevs, denoted $v$ and $x_0$ respectively, and the mass parameters may be written as 
\begin{align}
	\mu^2 = \lambda_0 v^2 + \frac{a_2}{2} x_0^2 + \frac{a_1}{2} x_0 
	\qquad {\rm and} \qquad
	b_2 = -b_4 x_0^2 - b_3 x_0 - \frac{a_2}{2} v^2 - \frac{a_1}{4} \frac{v^2}{x_0} \, .
\end{align}
Provided that $x_0 \neq 0$, the cubic terms $s^3$ and $s h^2$ help to generate a barrier separating the symmetric and broken vacua and make the PT strongly first order.  
A number of PT analyses \cite{Profumo:2007wc, Espinosa:2011ax, Ham:2004cf, Ahriche:2007jp} have revealed that the xSM can accommodate a strongly first order electroweak PT.  
They also find that this model displays multiple patterns of symmetry breaking such that, either $h$ and $s$ can obtain vevs at the same temperature, or $s$ can receive a vev prior to electroweak symmetry breaking.  
If we were to search for SFOPT by randomly choosing order one parameters, there would be no way of anticipating what pattern of symmetry breaking would be realized, or if the EW symmetry would be spontaneously broken at all.  
Moreover, since \eref{eq:xSM_U} has six free parameters, such a random search could become quite time consuming.  

The discrete symmetry technique greatly simplifies the SFOPT search.  
We are able to specify a desired pattern of symmetry breaking to investigate, identify the corresponding discrete symmetry, compute the associated EDSP, and begin searching by perturbing from the EDSP.  
Here, we will focus on a particular pattern of symmetry breaking in which both $s$ and $h$ obtain vevs simultaneously, and we will compare our calculation against the ``high-T trivial singlet vev'' case of \cite{Profumo:2007wc}.   
The appropriate discrete symmetry is a $\mathbb{Z}_2$ relating the vacua at $\left\{ h , s \right\} = \left\{ 0 , 0 \right\}$ and $\left\{ v , x_0 \right\}$.  
We can identify the associated EDSP by first reducing \eref{eq:xSM_U} to \eref{eq:GenSing_U} and then imposing $\alpha = 1$.  
This is accomplished by focusing on the one-dimensional linear trajectory $\left\{ h, s \right\} = \left\{ v, x_0 \right\} \times \varphi / \sqrt{v^2 + x_0^2}$ parametrized by $\varphi$, which interpolates between the EW-symmetric and EW-broken vacua.  
Along this trajectory, the potential can be written in the form of \eref{eq:GenSing_U} with 
\begin{align}\label{eq:xSM_mapto1D}
	\lambda & = \frac{\lambda_0 v^4 + b_4 x_0^4 + a_2 v^2 x_0^2}{(v^2 + x_0^2)^2} 
	&\qquad \qquad
	\mathcal{E} & = - \frac{x_0( 3 a_1 v^2 + 4 b_3 x_0^2)}{12 (v^2 + x_0^2)^{3/2}} \nonumber \\
	M^2 & = \sqrt{v^2+x_0^2} \left( 3 \mathcal{E} - \lambda \sqrt{v^2+x_0^2} \right) 
	&\qquad \qquad
	\alpha & = \frac{\lambda M^2}{2 \mathcal{E}^2} \, .
\end{align}
Then, upon resolving the condition $\alpha = 1$ we find the enhanced discrete symmetry point,
\begin{align}\label{eq:xSM_EDSP}
	{\rm EDSP:} & \quad
	0 = 12 a_2 v^2 x_0^2 + 3 a_1 v^2 x_0 + 4 b_3 x_0^3 + 12 b_4 x_0^4 + 6 \lambda_0 v^4 \nonumber \\
	\mathbb{Z}_2 \ &: \ 
	\left( \varphi - \frac{\mathcal{E}}{\lambda} \right) \to - \left( \varphi - \frac{\mathcal{E}}{\lambda} \right) \, .
\end{align}
In general, the PT will not occur along the trajectory parametrized by $\varphi$, but nevertheless this linear interpolation is useful for identifying the EDSP.

\begin{figure}[t]
\begin{center}
\includegraphics[width=0.60\textwidth]{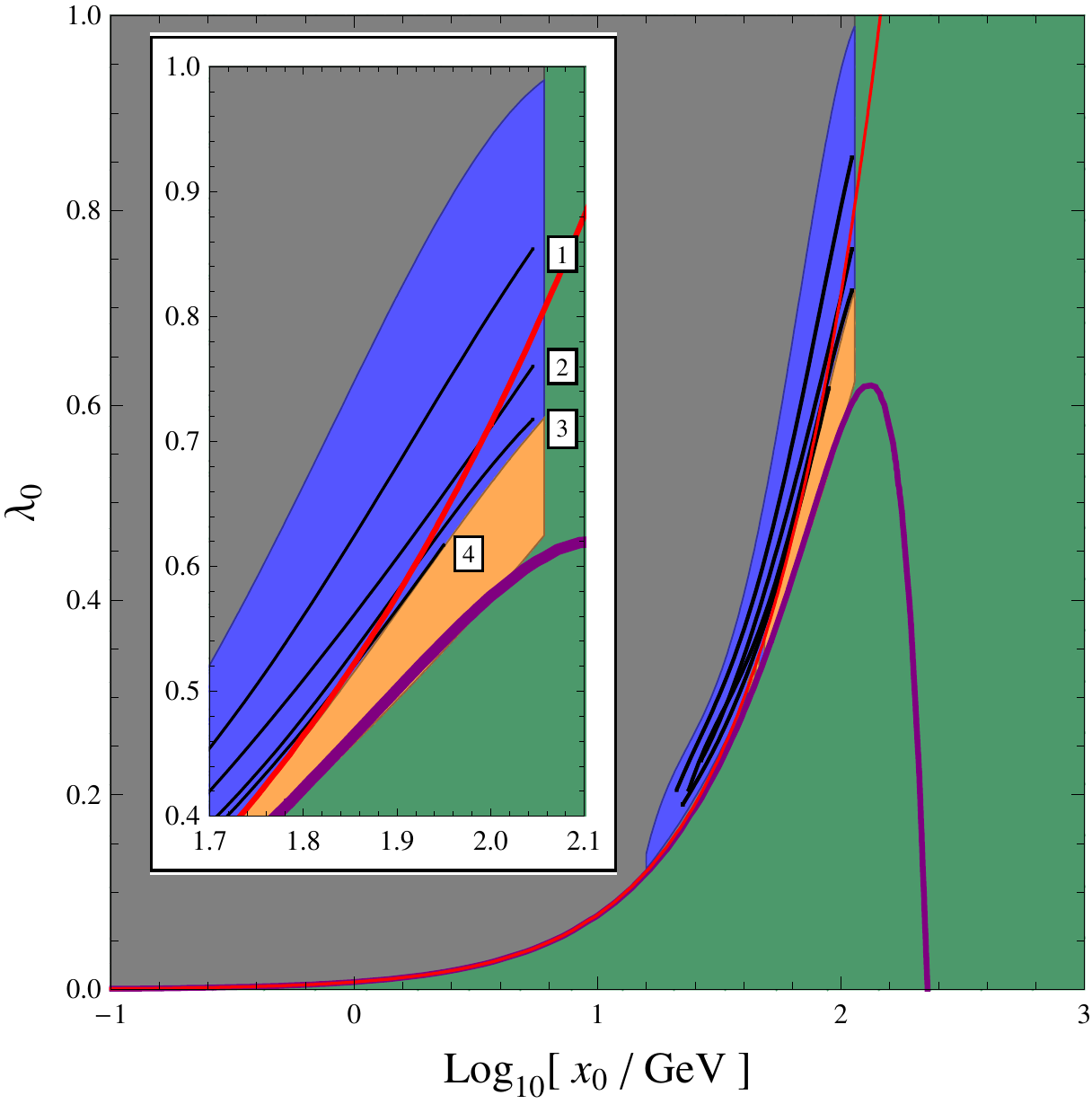}
\caption{\label{fig:xSM_scan1}  A slice of the xSM parameter space showing the proximity of SFOPT (\Csfopt \ region) to the enhanced symmetry axis (\Cesp \ curve).    
 }
\end{center}
\end{figure}

Once again, we have numerically investigated the strength of first
order PTs in the vicinity of the EDSP.  We have chosen a
parameter set which allows us to reproduce Fig.~4 (left panel) of
\cite{Profumo:2007wc} by fixing $a_1 = - 933 \GeV, a_2 = 0.69, b_3 =
356 \GeV, b_4 = 0.53$ and scanning $\lambda_0 \in \left[ 0, 1 \right]$
and $\log_{10} x_0 \in \left[ -1, 3 \right]$.  Our results are shown
in Fig.~\ref{fig:xSM_scan1}.  A few observations may be made.  First,
as anticipated, the first order PTs are strongest close
to the EDSP (\Cesp) curve and become weaker farther away.  
Second, there is a large region (\Cwronggm) in which the EW remains unbroken.  
Below the EDSP (\Cesp) curve, the origin remains the global minimum of the 
effective potential, whereas at large values of $x_0 \gtrsim 10^{2.1}$, the 
global minimum sits at $s < 0$.  
Third, in comparing with \cite{Profumo:2007wc}, one must bear in mind that we have fixed the remaining parameters, whereas those authors have scanned the full parameter space and projected onto these coordinates.  
As such, the region where we find SFOPT is much smaller than what is suggested by Fig.~4 of \cite{Profumo:2007wc}.
However, this just goes to show that it is typically difficult to find SFOPT in a large parameter space without either a large parameter scan or some guiding principle.

\vspace{1cm} 
\subsection{SM Plus Real $\mathbb{Z}_2$-Charged Singlet -- $\mathbb{Z}_2$xSM}\label{sub:Z2xSM}

As a final example, we turn out attention to the $\mathbb{Z}_2$xSM, which extends the SM by a real scalar singlet $s$ such that the scalar potential becomes \cite{Profumo:2010kp, Espinosa:2011ax}
\begin{align}\label{eq:Z2xSM_U}
	U(h, s) = \frac{\lambda}{4} h^4 - \frac{\mu^2}{2} h^2 + \frac{b_4}{4} s^4 + \frac{b_2}{2} s^2 + \frac{a_2}{4} \, s^2 h^2 \, .
\end{align}
The singlet is charged under a $\mathbb{Z}_2$, which restricts the allowed operators, but extends the possible patterns of symmetry breaking, because now $\left< s \right> = 0$ is radiatively stable.  
We will focus on a particular parameter region in which there is transitional $\mathbb{Z}_2$ symmetry breaking:  at temperature $T > T_a$ both $\mathbb{Z}_2$ and the EW symmetry are restored, at $T=T_a$ the singlet obtains a vev breaking $\mathbb{Z}_2$, and at $T=T_b < T_a$ the Higgs field obtains a vev and the singlet's vev returns to zero, thereby breaking the EW symmetry and restoring the $\mathbb{Z}_2$ (i.e., ${\rm EW} \times \mathbb{Z}_2 \to {\rm EW} \times \cancel{\mathbb{Z}_2} \to \cancel{\rm EW} \times \mathbb{Z}_2$).  
In the context of this pattern of symmetry breaking, the enhanced discrete symmetry point admits an $\mathbb{S}_2$ symmetry, 
\begin{align}
	{\rm EDSP:} \quad
	b_4 = \lambda \quad {\rm and} \quad b_2 = - \mu^2
	\qquad \qquad \qquad
	\mathbb{S}_2 \ : \ 
	h \leftrightarrow s \, 
\end{align}
where we will also take $a_2 > 2 \lambda$ to ensure that the discrete symmetry interchanges vacua.  
Note that this $\mathbb{S}_2$ symmetry is more restrictive than the $\mathbb{Z}_2$ symmetries we considered in the previous examples.  
To illuminate the role of the EDSP in locating SFOPT, we will reparametrize $b_4 = \lambda + \Delta b_4$ and $b_2 = - \mu^2 + \Delta b_2$ to write the potential as 
\begin{align}\label{eq:Z2xSM_U_deltas}
	U(h,s) = \left[ \frac{\lambda}{4} \left( h^4 + s^4 \right) - \frac{\lambda v^2}{2} \left( h^2 + s^2 \right) + \frac{a_2}{4} h^2 s^2 \right] + \left[ \frac{\Delta b_4}{4} s^4 + \frac{\Delta b_2}{2} s^2 \right]
\end{align}
where we have also used $\mu^2 = \lambda v^2$.  
In this parameterization, we expect to find SFOPT nearby to the EDSP at $\Delta b_4 = \Delta b_2 = 0$.  

\begin{figure}[h]
\begin{center}
\includegraphics[width=0.30\textwidth]{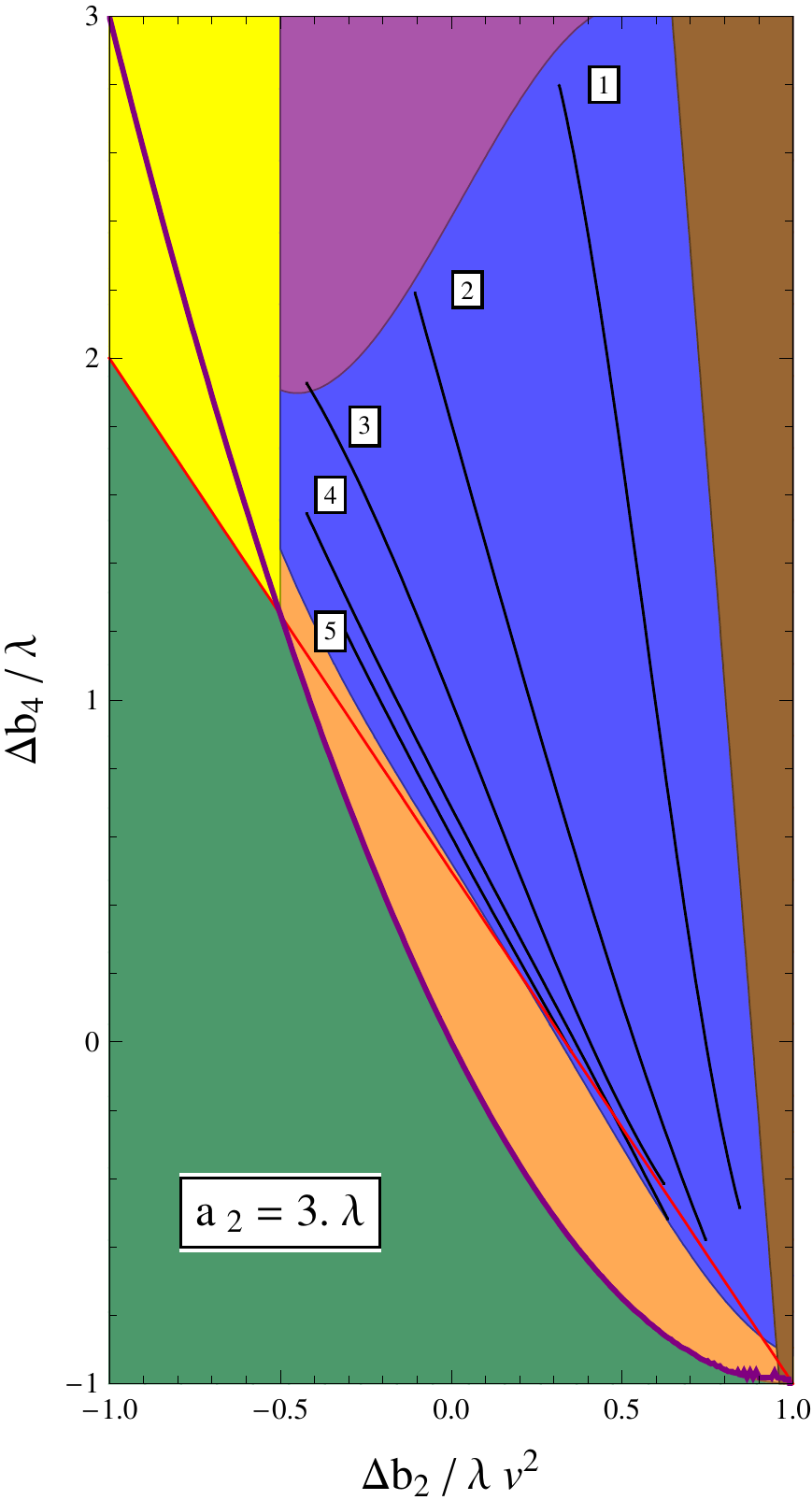} \hfill
\includegraphics[width=0.30\textwidth]{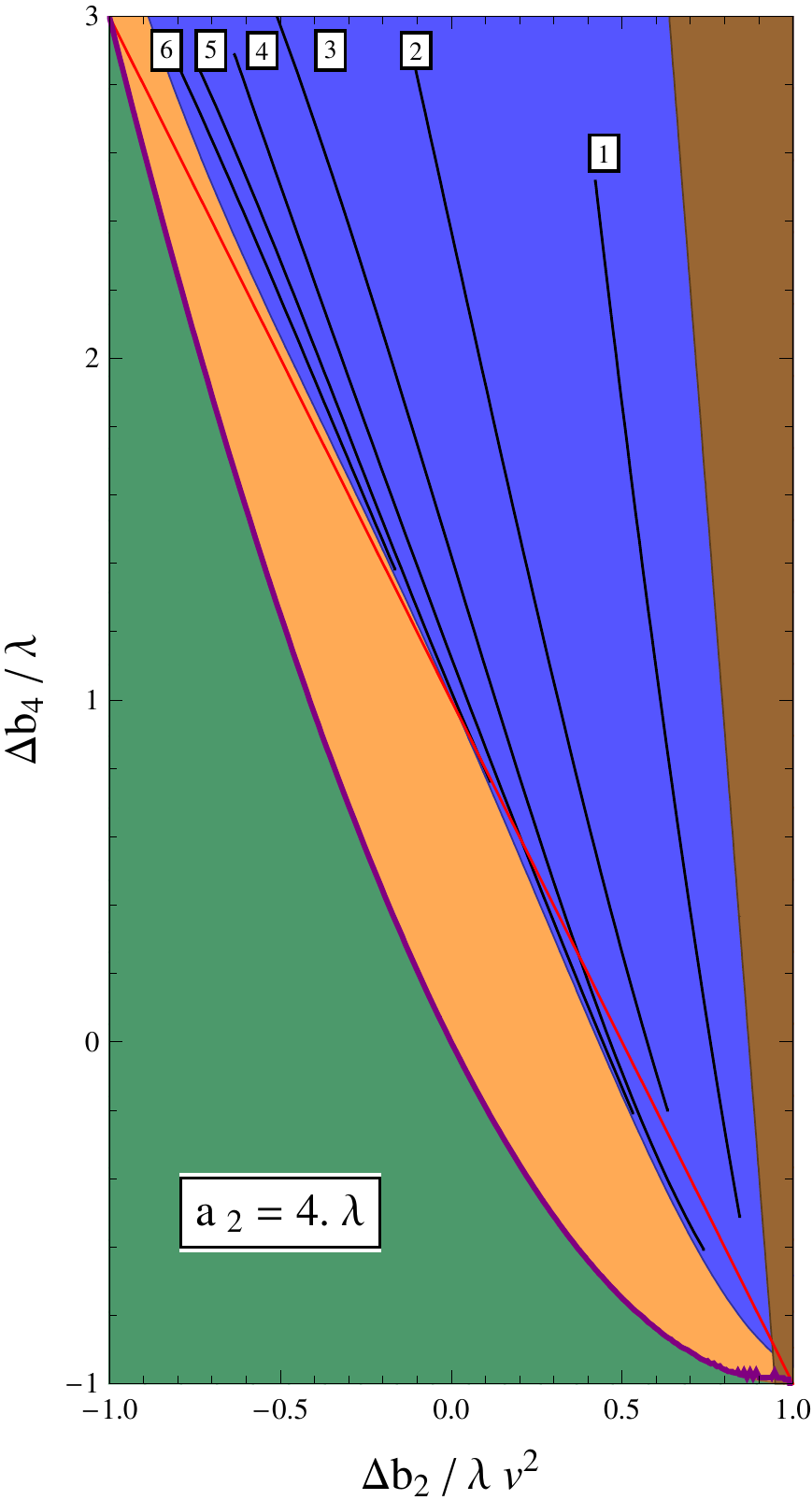} \hfill
\includegraphics[width=0.30\textwidth]{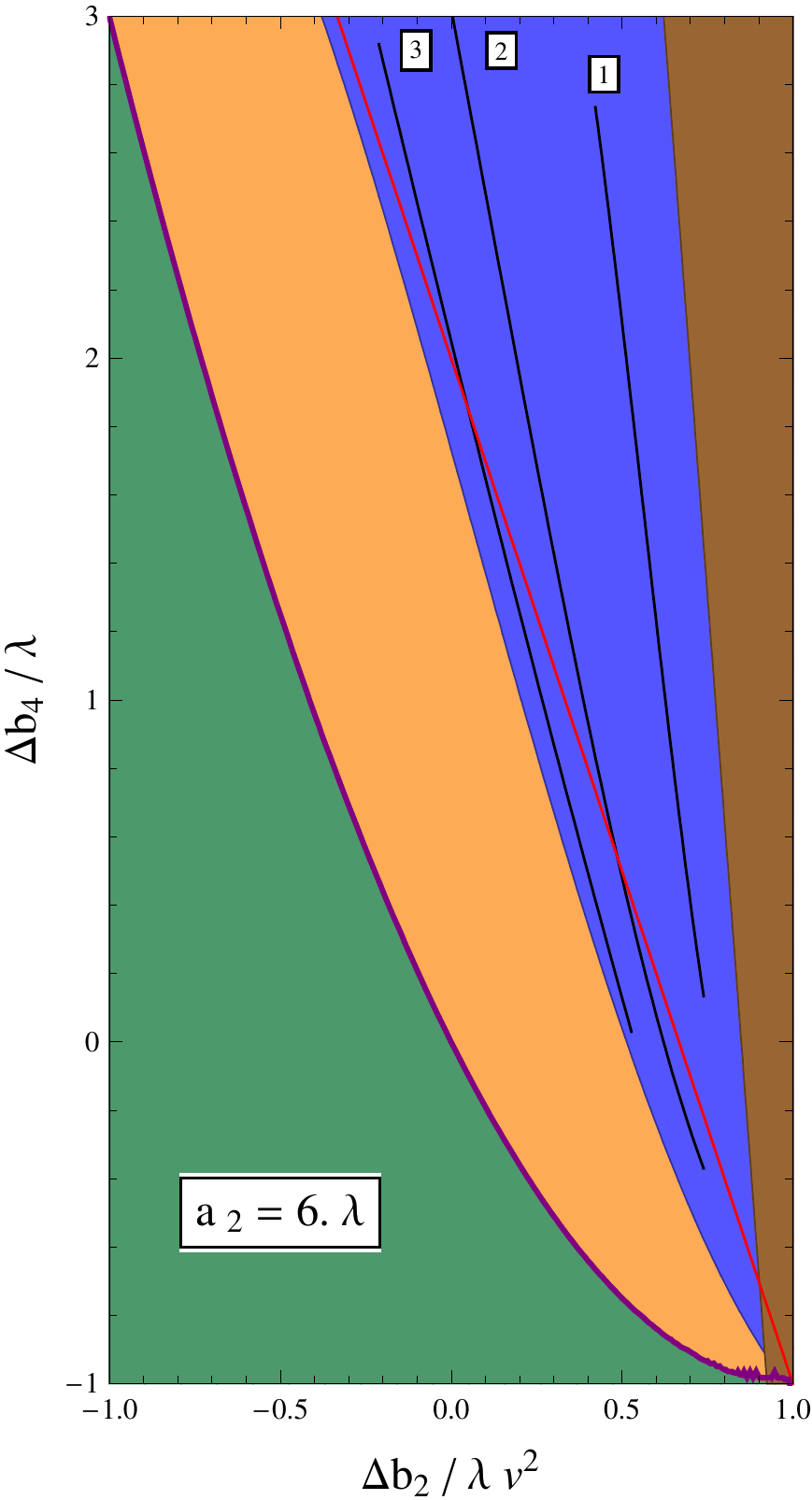} 
\caption{\label{fig:Z2xSM_scan1}  
Three slices of the $\mathbb{Z}_2$xSM parameter space for fixed $\lambda \approx 0.12$.  The origin $\Delta b_2 = \Delta b_4 = 0$ is an EDSP at which the theory has an $\mathbb{S}_2$ discrete symmetry.  
}
\end{center}
\end{figure}

We present the results of our numerical analysis in Fig.~\ref{fig:Z2xSM_scan1}, where we have fixed $\lambda \approx 0.12$ to give a Higgs mass\footnote{Since the axes of Fig.~\ref{fig:Z2xSM_scan1} depend only on the ratios $\Delta b_4 / \lambda$ and $\Delta b_2 / \lambda v^2$, a change in the Higgs mass (via $\lambda$) could be absorbed by $\Delta b_4$ and $\Delta b_2$, such that the qualitative features of Fig.~\ref{fig:Z2xSM_scan1} would remain unchanged.} of $m_h = \sqrt{ 2 \lambda v^2 } = 120 \GeV$.
As in the previous examples, the phase transition strength decreases monotonically with distance from the enhanced symmetry axis.
Significantly far from the EDSP, the phase transition proceeds with a different pattern of symmetry breaking.
In the brown region, the EW symmetry breaks without transitional $\mathbb{Z}_2$ violation (${\rm EW} \times \mathbb{Z}_2 \to \cancel{\rm EW} \times \mathbb{Z}_2$), in the yellow region the $\mathbb{Z}_2$ remains broken in the low temperature vacuum (${\rm EW} \times \mathbb{Z}_2 \to {\rm EW} \times \cancel{\mathbb{Z}_2} \to \cancel{\rm EW} \times \cancel{\mathbb{Z}_2}$), and in the purple region there exists an intermediate phase in which both $\mathbb{Z}_2$ and the electroweak symmetry are broken (${\rm EW} \times \mathbb{Z}_2 \to {\rm EW} \times \cancel{\mathbb{Z}_2} \to \cancel{\rm EW} \times \cancel{\mathbb{Z}_2} \to \cancel{\rm EW} \times \mathbb{Z}_2$). 

The region of parameter space nearby to the EDSP displays an interesting phenomenology.  
Since the singlet mass is given by 
\begin{align}
	m_s = m_h \left[ \frac{a_2/\lambda}{4} - \frac{1 - \Delta b_2 / \lambda v^2}{2} \right]^{1/2}
\end{align}
one typically finds $m_s \lesssim m_h$ nearby to the enhanced symmetry point.  
The unbroken $\mathbb{Z}_2$ symmetry ensures that the singlet is stable, and thus it is a dark matter candidate which annihilates to Higgses with a cross section proportional to $a_2^2$.  
A number of analyses \cite{Gonderinger:2009jp,Barger:2007im, McDonald:1993ex,Burgess:2000yq, He:2007tt,Davoudiasl:2004be, Bandyopadhyay:2010cc, Cohen:2011ec, Low:2011kp} have considered this scenario and found that $a_2$ and the singlet mass $m_s$ can be strongly constrained by assuming that the $s$ particle composes all of the dark matter.  
Collider experiments, such as the LHC, may also be able to constrain the Higgs-singlet coupling.  
For $\Delta b_2 / \lambda v^2 < (3 - a_2 / \lambda)/2$, the singlet mass is less than half of the Higgs mass and the invisible decay channel $h \to ss$ becomes kinematically accessible.  
Then, a measurement of the invisible decay width may constrain the Higgs-singlet coupling $a_2$ \cite{Englert:2011yb, Bock:2010nz, Barger:2003rs, Low:2011kp, Englert:2011er}.  
On the other hand, the singlet self-coupling $b_4$ remains unconstrained.  
This is because unlike in other limits of this model and similar models \cite{Noble:2007kk, Englert:2011yb, Bock:2010nz, Ashoorioon:2009nf}, the unbroken $\mathbb{Z}_2$ symmetry prevents the Higgs and singlet from mixing.  
Consequently, the singlet self-coupling $b_4$ is practically impossible to constrain at colliders, and contributions to the anomalous Higgs trilinear coupling \cite{Djouadi:1999rca} are loop suppressed.  
Finally, let us point out that the transitional $\mathbb{Z}_2$ violation limit may not suffer from the domain wall problem that generally accompanies models with spontaneously broken discrete symmetries.  
When the $\mathbb{Z}_2$ breaks in the first step of the PT, domain walls will be generated.  
However, once the EW symmetry is broken and the $\mathbb{Z}_2$ symmetry is restored, the domain walls should be ``wiped out'' by the $\mathbb{Z}_2$-symmetric vacuum field configuration.  
This may lead to a unique gravitational wave spectrum.


\section{\label{sec:conc}Conclusion}

Strongly first order phase transitions (SFOPTs) are required for electroweak baryogenesis and may have other interesting implications for early universe relics.  
In this article we have discussed a general analytic guideline, based on symmetry principles, which is useful in identifying a region of parameter space favorable for SFOPT: an arbitrarily strong PT can be found for parameters near an enhanced discrete symmetry point (EDSP) if the condition for spontaneous symmetry breaking is met and  if the discrete symmetry relates the electroweak symmetry preserving vacuum to one in which it is broken. 
Group theoretically, this means that the coset representation of the broken discrete symmetry contains an electroweak singlet and the discrete group does not commute with the electroweak group.
Because of phenomenological requirement of completing the PT at a nonzero temperature, the symmetry must be broken by parametric deformations away from the EDSP.  
As the deformation decreases, the strength of the PT tends to increase.  
We applied this guideline to study the electroweak PT in three specific models.  
In each of the models considered, SFOPTs occur in close proximity to the EDSP, as expected.  
In this way, the enhanced symmetry point acts like a lamppost in the parameter space, signaling the location of SFOPTs.  
It would be interesting to apply a similar EDSP-motivated analysis of the electroweak phase transition to models with larger scalar sectors and greater parametric freedom, such as singlet extensions of the Minimal Supersymmetric Standard Model.  

It is not unnatural to expect SFOPT to be localized in the vicinity of an EDSP.  
Strongly first order phase transitions almost always require some fine-tuning of the parameters in the theory.  
From an UV completion point of view, such fine-tuning could be more natural if it is close to a point of the parameter space with enhanced symmetry. 
It is also clear that degenerate vacua may be found even without discrete symmetry, and thus our guideline provides a sufficient, though not necessary, condition for locating SFOPT.  
Nonetheless, such parametric regions form a large class of possibilities which can most likely always occur in practice. 

We also observe (as did \cite{Espinosa:2011ax, Chung:2011hv}) that the PT tends not to proceed at all unless the barrier separating the EW-broken and EW-unbroken vacua is very small or not present at all (along the \Ctachyon \ curve), because otherwise the tunneling rate is too strongly suppressed.
Hence, the deformations away from the EDSP required for phenomenologically viable SFOPTs are not vanishingly small and are model dependent.  
Although such phenomenologically viable parametric regions can be arrived at by deforming away from enhanced continuous symmetry points rather than EDSPs, the EDSP starting point guarantees the existence of potential barriers required for a first order PT.  
In that sense, our proposal here is advantageous over the enhanced continuous symmetry point perspective.

Proximity to an EDSP implies interesting relations between parameters in the extended Higgs sector, which is responsible for the dynamics of the electroweak symmetry breaking. 
Such relations will manifest themselves in both the spectrum of the states in the Higgs sector and their couplings.  
Probing this sector is the central scientific focus of the LHC.  
We might have already seen the discovery of the Higgs boson on the horizon \cite{AtlasHiggs,CMSHiggs}.  
Discovering the additional states in the extended Higgs sector and measuring the parameters in the Higgs potential are expected to be very challenging tasks.  
At the same time, confirming the structure of the Higgs sector to be consistent with a SFOPT would establish a striking link to the generation of the baryonic asymmetry in the universe.


\begin{appendix}

\section{Details of Phase Transition Calculation}\label{app:tempdefs}

For the phase transition analyses in this paper, we have calculated the thermal effective potential $V_{\rm eff}(\vec{\phi}, T)$ through one-loop order using the standard techniques \cite{Coleman:1973jx, Dolan:1973qd, Jackiw:1974cv}.  
We numerically minimize\footnote{This definition of $v(T)$ implies that $T_c, T_n,$ and $T_r$ will be dependent upon the choice of gauge \cite{Patel:2011th, Loinaz:1997td}.  Though this may affect the numerical accuracy of our results, we expect that the qualitative parametric dependence of the EW order parameter nearby to an EDSP, which is our primary interest, will remain unchanged.  } $V_{\rm eff}$ with respect to $\vec{\phi}$ to obtain the scalar field expectation values in the symmetric and broken phases, $\vec{v}_{\rm sym}(T)$ and $\vec{v}_{\rm brk}(T)$, respectively.  
The latter quantity is sometimes referred to in the text as simply $v(T)$.  
The critical temperature $T_c$ is defined as 
\begin{align}
	V_{\rm eff}(\vec{v}_{\rm sym}(T_c), T_c) = V_{\rm eff}(\vec{v}_{\rm brk}(T_c), T_c) \, .
\end{align}
We use $V_{\rm eff}(\vec{\phi}, T)$ to calculate the action\footnote{For the models of Secs. \ref{sub:xSM} and \ref{sub:Z2xSM} which have more than one scalar field participating the phase transition, we calculate the bounce using the approximation described in \cite{Chung:2011it}.} $S_3(T)$ of the bubble field configuration that mediates the vacuum transition \cite{Linde:1977mm, Linde:1981zj, Coleman:1977py, Callan:1977pt}.  
We determine the bubble nucleation temperature $T_n$ by requiring the bubble nucleation rate per Hubble volume to exceed the Hubble expansion rate.  
This condition may be resolved to 
\begin{align}
	S_3(T_n) / T_n = 140
\end{align}
where the value on the right hand side depends only logarithmically on the model parameters \cite{McLerran:1990zh, Anderson:1991zb}.  
Finally, we calculate the temperature $T_r$ of the plasma after the phase transition ends and the plasma has been reheated.  
This is obtained by assuming that the universe does not expand significantly during the phase transition and then by imposing energy conservation \cite{Chung:2011it}
\begin{align}
	\rho_{\rm sym}(T_n) = \rho_{\rm brk}(T_r)
\end{align}
where 
\begin{align}
	\rho(T) = V_{\rm eff}(v(T),T) - T \frac{d}{dT} V_{\rm eff}(v(T),T)
\end{align}
is the energy density in the symmetric or broken phase, respectively.  

\end{appendix}
\begin{acknowledgments}

DJHC thanks Erik Weinberg for an interesting discussion regarding this
work at KIAS.  Preliminary results of this work were presented by DJHC
at ``Out-of-Equilibrium Quantum Fields in the Early Universe'' in
Aachen, September, 2010.  DJHC, VB, and AJL were supported in part by the
DOE through grant DE-FG02-95ER40896.  LTW is supported by the NSF
under grant PHY-0756966 and the DOE Early Career Award under grant
DE-SC0003930.

\end{acknowledgments}


\bibliographystyle{JHEP}
\bibliography{refs_DSymandPT}

\end{document}